\documentclass[aps,prl,showpacs,superscriptaddress,twocolumn,floatfix]{revtex4}
\usepackage{latexsym}
\usepackage{amsfonts}
\usepackage{amssymb}
\usepackage{amsmath}
\usepackage{graphicx}

\begin{document}

\title{Critical velocity for the vortex core reversal in perpendicular bias magnetic field}

\author{Alexey V. Khvalkovskiy}
\altaffiliation{Corresponding author. Electronic address: khvalkov@fpl.gpi.ru}
\affiliation{A.M. Prokhorov General Physics Institute of RAS, Vavilova str. 38, 119991 Moscow, Russia}
\affiliation{Unit\'e Mixte de Physique CNRS/Thales and Universit\'e Paris Sud 11, RD 128, 91767 Palaiseau, France}
\author{Andrei N. Slavin}
\affiliation{Oakland University, Rochester, MI-48309, USA}
\author{Julie Grollier}
\affiliation{Unit\'e Mixte de Physique CNRS/Thales and Universit\'e Paris Sud 11, RD 128, 91767 Palaiseau, France}
\author{Konstantin A. Zvezdin}
\affiliation{A.M. Prokhorov General Physics Institute of RAS, Vavilova str. 38, 119991 Moscow, Russia}
\affiliation{Istituto P.M. s.r.l., via Cernaia 24, 10122 Torino, Italy}
\author{Konstantin Yu. Guslienko}
\affiliation{Dpto. Fisica de Materiales, Universidad del Pais Vasco, 20018 San Sebastian, Spain}
\affiliation{IKERBASQUE, the Basque Foundation for Science, 48011 Bilbao, Spain}

\begin{abstract}

For a circular magnetic nanodot in a vortex ground state we study how the critical velocity $v_c$ of the vortex core reversal depends on the magnitude $H$ of a bias magnetic field applied perpendicularly to the dot plane. We find that, similarly to the case $H$ = 0, the critical velocity does not depend on the size of the dot. The critical velocity is dramatically reduced when the negative (i.e. opposite to the vortex core direction) bias field approaches the value, at which a \emph{static} core reversal takes place. A simple analytical model shows good agreement with our numerical result.

\end{abstract}

\maketitle

A magnetic vortex is a curling magnetization distribution in flat
magnetic submicron dots, with the magnetization pointing
perpendicularly to the dot plane within the ten nanometer size
vortex core. The vortex ground state corresponds to a deep energy
minimum when the dot lateral sizes fit the conditions for vortex
stability \cite{DingPRL2005}. This unique magnetic object has
attracted much attention recently because of the fundamental
interest to specific properties of such a nanoscale spin structure.
The direction of the core polarization ('up' or 'down') can store a
bit of information, and this is of considerable practical interest
for applications in magnetic memory technology. Several
approaches can be used to switch the bit, i.e., to reverse the
vortex core. It has been shown that a static magnetic field can
reverse the core if its magnitude reaches sufficiently large values,
typically of several kOe \cite{Okuno, ThiavillePRB03}. The reason
for that is a large energy barrier between the vortex states with
'up' or 'down' core polarizations. Alternately, the magnetic core
can be switched at zero static magnetic field, if it is excited by a
variable (oscillating or pulsed) in-plane field or by a
spin-polarized current \cite{VanWaeyenberge, Choi, Hertel, Yamada}.
The reversal occurs if the core velocity reaches a certain critical
value $v_c$, which is defined solely by the magnetic parameters
\cite{GuslienkoPRL08}. Very recently, the vortex core switching has
been observed at intermediate experimental conditions in Ref.
\cite{GdeLoubens}. In this work, a static perpendicular magnetic
field together with a small oscillating in-plane field was applied
to a nanodot in a vortex state. The frequency of the excitation was
swept and the resonant vortex motion was detected. At a given
magnitude of the exciting field, the vortex core was switched when
the static bias magnetic field reached a critical value.

In our numerical study, we calculate the critical velocity of vortex core reversal as a
function of a static out-of-the plane magnetic field. We find that
this critical velocity, similarly to the case of zero applied field,
is  independent of the dot sizes, but depends on the magnetic
parameters of the dot material. The critical velocity drops
significantly with the increase of the magnitude of the negative
(opposite to the vortex core direction) bias magnetic field.

\begin{figure}[h]
   \centering
    {\includegraphics[keepaspectratio=1,width=8.5 cm]{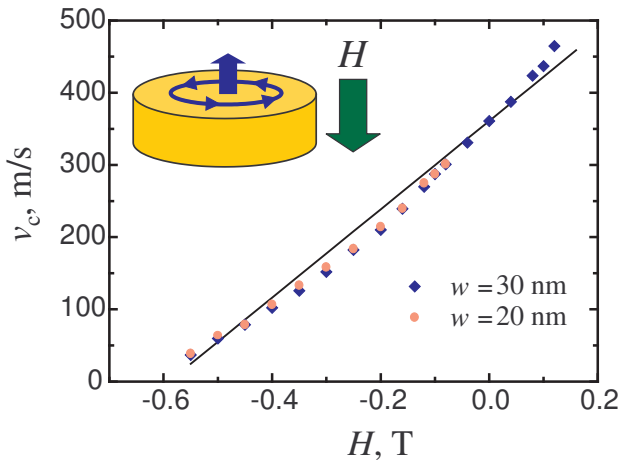}}
     \caption{(color online) Symbols: critical velocity $v_c$ as a function
    of the magnitude of the magnetic field applied perpendicularly to the dot
    plane for dots with thickness $w$ = 20 nm and $w$ = 30 nm.
    Solid line: analytical prediction by Eq. \ref{AnalytVc}}
\label{Fig1}
\end{figure}

We consider a circular magnetic  nanodot in a vortex ground state. A
static bias magnetic field is applied perpendicularly to the dot
plane (along  the $z$-axis). It is considered to be positive if
parallel (and negative if antiparallel) to the initial direction of
the vortex core (or vortex core polarization), see the inset to Fig.
\ref{Fig1}. The vortex motion is excited by a d.c. spin polarized
current flowing perpendicular to the dot plane \cite{KhvalkovPRB}.
The spin polarization of the current is along the $z$-axis. The vortex motion is calculated numerically
using our micromagnetic code \cite{NoteCalculations}. Two dots with a diameter 2$R$ = 300
nm and thicknesses $w$ = 20 nm and 30 nm were considered. Magnetic
parameters mimic those for NiMnSb used in Ref. \cite{GdeLoubens}:
the saturation magnetization is $4 \pi M_s$ = 0.69 T, the easy plane
anisotropy field $H_A$ = 0.185 T, the exchange stiffness A = 1
$\times$ 10$^{-11}$ J/m and the Gilbert damping $\alpha$ = 0.01. The
mesh cell size is 1.5 $\times$1.5 $\times$ 5 nm$^3$.

Due to the excitation by the spin  current, the vortex core starts
to gyrate with gradually increasing radius. Correspondingly, the
core velocity is increasing until it eventually switches. Prior to
the core reversal, a region with negative values of $M_z$ component
(a 'magnetic dip') is formed at the inner part of the core
trajectory. When the core velocity reaches the critical value, this
dip splits into a vortex with negative polarization and an
antivortex \cite{Hertel}. The antivortex annihilates with the
original vortex core, and in the end only the vortex with a negative
polarization remains. After the core reversal the spin-polarized
current starts to damp the core gyration, thus slowing the core
motion and, eventually, bringing the reversed core to the
equilibrium position in the dot center \cite{KhvalkovPRB}.

The critical velocity $v_c$ is determined as the maximum core
velocity on the gyration trajectory. This maximum is reached just
before the core switching. The core velocity is calculated as the
time derivative of the core position, which in its turn is extracted
from the magnetization distributions printed out each 0.05 ns. At
fields larger than 0.12 T for the dot with $w$ = 30 nm
(correspondingly, larger than $-0.04$ T for the dot with $w$ = 20 nm)
the core is expelled from the dot prior to the switching. At fields
smaller than $-0.55$ T the mesh we use becomes insufficiently fine
to be able to calculate the vortex dynamics accurately. The static
switching field $H_c$ is equal to $-5.9$ kOe for the both dot
thicknesses \cite{NoteHstatic}. This approximate numerical value for
$H_c$ was used below in our analytical calculations of the
dependence of the critical velocity on the perpendicular bias field.
The results of this calculation are shown by a solid line in Fig. 1.

The simulation results for the two dots are summarized in Fig.
\ref{Fig1} (symbols). The critical velocity at $H$ = 0 is $v_c$ = 360
m/s for both dots. $v_c(H)$ increases for increasing positive fields
($v_c$ = 460 m/s for $w$ = 30 nm at $H$ = 0.12 T). However it
diminishes significantly for negative fields ($v_c$ = 40 m/s at H =
$-$ 0.55 T for both the dots). At moderate fields ($\left| H
\right|$ $<$ 2 kOe) $v_c$ scales linearly with H, and the slope is
approximately 670 $m (s T)^{-1}$.

We find that for all the field values, when they can be calculated,
the critical velocities determined for the two dots coincide.  This
fact is rather nontrivial as, owing to different thicknesses, many
parameters of the vortex (such as the profile of the potential
energy $W(\mathbf{X})$, where $\mathbf{X}$ is the core position
\cite{GuslienkoJAP2002}; the separation of the vortex from the dot
center and edges at the switching; the shape of the core) are very
different for the two dots and depend differently on the field. From
these results we conclude that, similarly to the case $H$ = 0, for
non-zero $H$, $v_c$ depends only on local properties of the vortex
core spin structure \cite{NoteWhyLocal}.

The zero-field value of the critical velocity for both dot
thicknesses, $v_c$ = 360 m/s, is in perfect agreement with the
analytical prediction of the works of Ref. \cite{GuslienkoPRL08,
KSLee} taking into account the easy-plane anisotropy constant
$K=M_sH_A/2$ of NiMnSb: $v_c(0)$ = $1.66 M_s \sqrt{2\pi A / (2\pi
M_s^2 + K)}$, which gives $v_c$ = 340 m/s.  In the following, we
investigate the underlying physics responsible for the $v_c$(H)
behavior presented in Fig. 1. The vortex core dynamic reversal, as
it was shown in Ref. \cite{GuslienkoPRL08}, originates from the
self-induced dynamic gyrotropic field or gyrofield. This field is
induced by the vortex motion and its amplitude is proportional to
the ratio $v/\rho$, where $v$ is the velocity of the moving vortex
and $\rho$ is the vortex core radius. When the gyrofield reaches a
critical value $H_{g}^{cr}$ $\propto v_c/\rho$, the vortex core very
rapidly reverses.

\begin{figure}
   \centering
    \includegraphics[keepaspectratio=1,width=8.5 cm]{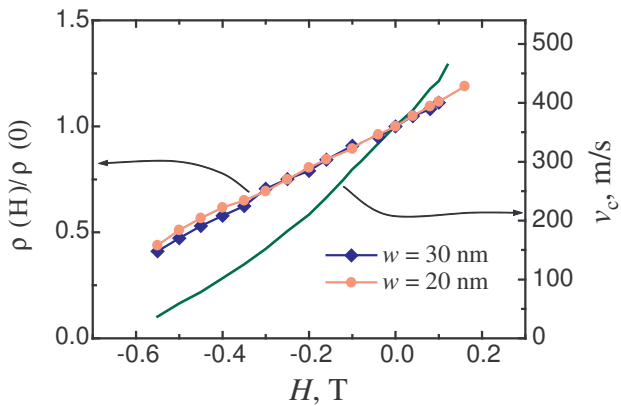}
     \caption{(color online) Symbols: radius of the vortex core $\rho$ as a function of magnetic field $H$, for two dots. Solid line: critical velocity for the dot with $w$ = 30 nm.}
    \label{Fig3}
\end{figure}

We study how $\rho$ scales with $H$ for the two dot thicknesses.  We
analyze magnetization distribution profiles for a static vortex in
equilibrium at different fields to extract the dependence $\rho (H)$
\cite{NoteHowWeGetRc}. We find that although $\rho (H)$ is different
for the two dots  (e.g., at $H$ = 0, $\rho$ = 20 nm for the dot with
$w$ = 30 nm and correspondingly $\rho$ = 18 nm for $w$ = 20 nm), to
the precision of our calculation, $\rho (H)/\rho(0)$ coincides for
the two thicknesses, as can be seen on Fig. \ref{Fig3}. From this
result we conclude that the critical value of the gyrofield
$H_{g}^{cr}$ $\propto v_c/\rho$ scales equally with the field $H$
for the two dots. It also indicates that at non-zero external field
the critical velocity $v_c (H)$ relies on the same vortex core
reversal mechanism than at zero field, i.e. it is mainly determined
by a competition of the gyrotropic and exchange fields within the
core. The gyrofield deforms the core magnetization profile, whereas
the exchange field tries to create a more uniform magnetization
distribution suppressing the core deformation.

As can be seen from Fig. \ref{Fig3}, the slopes of $v_c$ and $\rho$
as functions of the perpendicular magnetic field are different;
indeed, $v_c(H)$ decreases noticeably more rapidly than $\rho(H)$ at
negative $H$. This means that the critical value of the gyrofield
$H_{g}^{cr}$ decreases with negative $H$. This feature can be
attributed to the fact that the deformation of the core by the
perpendicular bias field $H$ leads to a decrease of the
effective potential barrier that the gyrofield has to surpass to
induce the vortex core reversal. Therefore, the bias field
provides two different actions on the vortex which help to switch
the core. One is that, at a given core velocity $v$, the amplitude of the gyrofield $H_{g}$ increases with negative $H$ as long as $\rho$ is reduced. Second is that the critical value of the
gyrofield $H_{g}^{cr}$ that is required to switch the core becomes smaller at higher negative fields. For any
field $H$, the vortex core reversal mode is an axially
\textit{asymmetric} mode like it was found for H = 0 \cite{Choi,
Hertel}. That is very different from the axially symmetric reversal
path involving the Bloch point (BP) found in the simulations for
static reversal \cite{ThiavillePRB03}. We also see this axially
symmetric mode and the BP formation in our simulations of the static
core reversal. But this axially symmetric BP mechanism is an
idealization, which leads to higher values of $H_c$. It can be not
realized practically due to unavoidable spontaneous symmetry
breaking in real systems, e.g. induced by the thermal fluctuations.
This is the vortex gyrotropic mode with a finite $\mathbf{X}$ that
is responsible for the axial symmetry breaking. That is why the
critical velocity $v_c(H)$ of the moving vortex is important.

We can get a simple analytical expression for the $v_c(H)$. Let us
consider a dot with static switching field $H_c$. The value of the
core radius at this field $\rho_c(H_c)$ is finite. The physical
sense of $\rho_c(H_c)$ is the following: the vortex with positive
polarization  becomes unstable in the point $H=H_c$ when decreasing
$H$. From the other side, it is reasonable to assume that the
dependence $v_c(H)$ goes to $0$ when $H$ approaches the static core
reversal field $H_c$; i.e., we can assume that $v_c(H)$ is
proportional to $(1 - H/H_c)$ near $H_c$. That immediately leads to
the dependence:
\begin{equation}\label{AnalytVc}
v_c(H) = v_c(0) (1 - H/H_c),
\end{equation}
The static field reversal and dynamic reversal mechanisms help each
other leading to descending dependence of $v_c(H)$. Thus, the
analytically estimated slope of the dependence $v_c(H)$ is $dv_c/dH$
= $-v_c(0)/H_{c}$ = 610 m/s T (shown as a solid line in Fig. 1),
that is close to the numerically simulated slope of 670 m/s T. These
speculations explain the main features of our simulations of
$v_c(H)$, $\rho_c(H)$ presented in Fig. 1 and 2.

In summary, our numerical study has demonstrated that there are two
contributions to the process of the vortex core reversal in a
magnetic dot subjected to a perpendicular bias magnetic field: the
static reversal mechanism related  to the instability of the vortex
core with polarization directed against the bias field and the
dynamical reversal mechanism related to the vortex core deformation.
While the first mechanism keeps the axial symmetry of the vortex
magnetization distribution, the second one breaks this axial
symmetry and creates an "easy" core reversal path. Thus, the
perpendicular bias magnetic field applied oppositely to the vortex
core direction reduces the critical velocity of the vortex core
reversal and facilitates the dynamical reversal process, which was
demonstrated experimentally in \cite{GdeLoubens}.

The work is supported by the EU project MASTER (grant 212257), RFBR
(grants 09-02-01423 and 08-02-90495), the  National Science
Foundation of the USA (grant No. ECCS 0653901), and by the U.S. Army
TARDEC, RDECOM (contract N0. W56HZW-09-P-L564). K.G. acknowledges
support by IKERBASQUE (the Basque Science Foundation).


\begin{thebibliography}{32}
\bibitem{NoteCalculations}
The code performs numerical integration of the LLG equation using the forth order Runge-Kutta method with an adaptive time-step control. The magnitude of the spin transfer term is $\left(\sigma J\right)$ = 6 mT in terms of Ref. \cite{KhvalkovPRB}. The Oersted field generated by the current is disregarded in the simulations.
\bibitem{NoteHowWeGetRc}
For a magnetization distribution in equilibrium, we consider the two-dimensional function $M_z(x,y)$ that is taken at one of two middle cell planes of the dot. $\rho$ at a given H is determined as FWHM of the function $M_z(x,y)$, given that the magnetization at sufficient separation from the vortex core defines the ground level, and the maximum is in the core center.
\end{thebibliography}
\end{document}